\documentclass[onecolumn,showpacs,preprintnumbers,superscriptaddress,amsmath,amssymb]{revtex4}
\usepackage{tipa}
\usepackage{amsmath}
\usepackage{txfonts}
\usepackage{amssymb}

\usepackage{graphicx}
\usepackage{dcolumn}
\usepackage{bm}
\usepackage{color}

\makeatletter 

\newcommand{\Rmnum}[1]{\expandafter\@slowromancap\romannumeral #1@}
\makeatother

\begin{document}

\large

\preprint{APS/123-QED}

\title{Structure and control of self-sustained target waves in excitable small-world networks}

\author{Yu Qian}
\affiliation{%
Department of Physics, Beijing Normal University, Beijing 100875, China
}%
\affiliation{%
Nonlinear Research Institute, Baoji University of Arts and Sciences, Baoji 721007, China
}%
\author{Xiaodong Huang}
\affiliation{%
Department of Physics, Beijing Normal University, Beijing 100875, China
}%
\author{Gang Hu}%
\email{ganghu@bnu.edu.cn}
\affiliation{%
Department of Physics, Beijing Normal University, Beijing 100875, China
}%
\author{Xuhong Liao}
\email{liaoxuhong@mail.bnu.edu.cn}
\affiliation{%
Department of Physics, Beijing Normal University, Beijing 100875, China
}%

\date{\today}

\begin{abstract}
  Small-world networks describe many important practical systems
among which neural networks consisting of excitable nodes are the
most typical ones. In this paper we study self-sustained
oscillations of target waves in excitable small-world networks. A
novel dominant phase-advanced driving (DPAD) method, which is
generally applicable for analyzing all oscillatory complex networks
consisting of nonoscillatory nodes, is proposed to reveal the
self-organized structures supporting this type of oscillations. The
DPAD method explicitly explores the oscillation sources and wave
propagation paths of the systems, which are otherwise deeply hidden
in the complicated patterns of randomly distributed target groups.
Based on the understanding of the self-organized structure, the
oscillatory patterns can be controlled with extremely high
efficiency.
\end{abstract}

\maketitle
\section{Introduction}
  Pattern dynamics in excitable media has attracted great attention
in wide fields due to its relevance to various important systems,
such as cardiac tissues and neural systems for typical examples
\cite{Winfree1987, Gray1998, Vogels2005}. Though single excitable
cell is not oscillatory, organized oscillatory patterns, however,
are extremely important issues in the media of coupled excitable
cells \cite{Bazhenov1999, Steriade2003, Buzsaki2004}. Spiral waves
and target waves are two typical patterns in excitable tissues
\cite{Winfree1987, Mikhailov1994}. The former can self-sustain in
autonomous systems with the spiral tips serving as the oscillation
sources while the latter can exist only by external pacing and can
never exist in autonomous regular excitable media. So far most of
researches studying pattern dynamics of excitable tissues have
focused on locally and regularly coupled media, and little has been
known on the influences of long-range or random links on the system
dynamics. In many realistic systems of great importance, of which
neural networks are typical examples, these long-range couplings do
exist and they play crucial roles in deciding the functions of the
systems \cite{Watts1998, Barabasi1999, Eguiluz2005, Qi2003,
Zumdieck2004, Lago2000, He2002, Yonker2006}. Recently, A. Roxin
\emph{et al.} \cite{Roxin2004}, S. Sinha \emph{et al.}
\cite{Sinha2007}, A. J. Steele \emph{et al.} \cite{Steele2006} and
Gray \emph{et al.} \cite{Gray1996} studied excitable systems in
small-world networks and found some new types of self-sustained
oscillations, including target waves in autonomous systems[18,19].
However, the mechanism underlying these new types of oscillations
and the effective methods to control and regulate the oscillatory
dynamics have far from been clear.

  In this paper we study the oscillatory behaviors of self-sustained
target waves in small-world networks of excitable nodes. We proposed
a novel method of \emph{dominant phase-advanced driving} to reveal
the successive driving structures supporting the self-sustained
oscillations, based on the known network structure and oscillation
data. In these structures oscillation sources and wave propagation
pathways are explored explicitly, which are otherwise deeply hidden
in randomly distributed target groups. Based on the structure we are
able to effectively control and regulate the oscillations of the
excitable networks by suitably manipulating very few long-range
links.

  The paper is arranged as follows. In Sec. \Rmnum{2} we describe
the basic idea and the operating procedure of dominant
phase-advanced driving (DPAD) method. Then we explain based on the
graph theory why this method can reveal the essential structure
generating self-sustained oscillations from complicated oscillatory
patterns of complex networks. In Sec. \Rmnum{3} we study a model of
small-world networks consisting of excitable nodes. We reveal both
oscillatory patterns of spiral waves and multiple-target waves. The
mechanism of the former case is well known \--- spiral wave tips
serve as the oscillation generators. In the latter case, we apply
the DPAD method and explicitly explore the wave source and the wave
propagation pathways. In Sec. \Rmnum{4} we show how to effectively
control self-sustained target patterns based on the above
understanding. In Sec. \Rmnum{5} we analyze the conditions for the
applications of the DPAD method, and show how this method can
provide useful understanding even if partial necessary information
are not available. The last section gives brief discussion on the
results and the significance of our method.

\section{Analyzing self-sustained oscillation of complex networks by using DPAD method}

  Considering a general network graph $G(V,E)$ with $V$ representing a set of
$N$ nodes and $E$ being a set of $M$ interactions (i.e., couplings).
Dynamic variables are associated to each node, and these variables
obey well defined coupled ordinary differential equations (ODEs).
Each node is nonoscillatory individually while the entire complex
networks are periodically oscillatory. It is well known that a
necessary condition for this type of oscillatory networks is that
there must exist some interacting loops. For any connected network
with $N$ nodes and $M$ interactions, the number of fundamental
cycles is $M-N+1$ \cite{Bollobas1998}. And the number of topological
cycles serving as the candidates of source loops are $2^{M-N+1}-1$,
which is huge in the case $M$ $\gg$ $N$ (this is so for most of
practical networks). Cycles are referred to as loops in our paper
for simplicity. In this section we will propose an effective method
to identify the key dynamic loops generating the oscillations from
the large number of topological loops.

  Regardless of different dynamics and different coupling forms,
we propose a common design principle for such oscillatory networks.

  Design principle: each nonoscillatory node can oscillate if and only
if it is driven by one or few oscillatory interactions with
\emph{advanced phases}.

\begin{figure}
\resizebox{14.0cm}{10.0cm}{\includegraphics{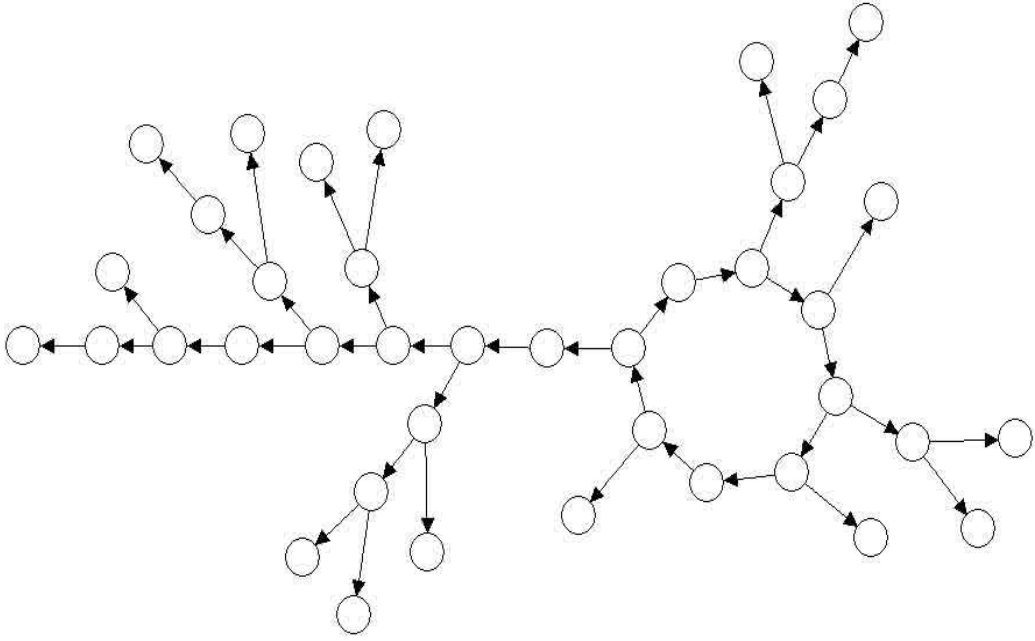}}\quad
\caption{\label{fig1} Schematic figure of universal structure of
periodically oscillatory 1D network consisting of nonoscillatory
nodes. The figure has a form of tree branches radiating from an
interacting loop, and is called tree from loop (TFL) structure.}
\end{figure}

  The definitions of $''$advanced phase$''$for different systems can be
different, but they are the same in essence\cite{Phaseadvanced}. Let
us consider an example of simplest $1D$ oscillatory networks
consisting of nonoscillatory nodes where each node is
unidirectionally interacted by only a single other node as shown in
Fig.~\ref{fig1}. The network consists of $N$ nodes with $M$
($M=N$)unidirectional interactions. As the network is oscillatory,
all the noes are activated. Suppose an arbitrary node $i_1$ is
phase-advancedly driven by a node $i_2$ via coupling, which is
phase-advancedly driven by node $i_3$ in turn, and this successive
unidirectional driving chain goes as $i_1\leftarrow i_2 \leftarrow
i_3 \leftarrow \cdots \leftarrow i_k \leftarrow \cdots$. Since $N$
is finite we must come to a node $i_q$, $q \leq N$, which is driven
by one of the previous nodes \{$i_1$, $i_2$, $\cdots$, $i_{q-1}$\},
said $i_p$ $(p<q)$. Then a successive regulatory loop $i_p
\leftarrow i_{p+1} \leftarrow \cdots \leftarrow i_q \leftarrow i_p$
is formed, serving as the oscillation source of all other nodes.
From the graph theory, this network has one and only one fundamental
loop ($N-N+1=1$), and this fundamental loop is right the dynamic
driving loop playing the role of the oscillation source, while all
other nodes must in the tree branches radiating from this loop
identifying wave propagation paths. The simple structure in
Fig.~\ref{fig1} consists of trees from loops, thus is called trees
from loops (TFL) pattern.

  The TFL structure of Fig.~\ref{fig1} is universal for all
self-sustained periodic oscillations in complex networks consisting
of nonoscillatory nodes. It illustrates the simplest relationship of
these nodes in self-sustained oscillation. Since no nonoscillatory
node can oscillate without phase-advanced driving from other nodes,
two key rules must be obeyed by any TFL structure:

  (i) There must be some (at least one) successively phase-advanced
  driving loops.

  (ii) Each node not in the loops must be in a tree branch rooted at
  a node in a loop.

  The simple and instructive structure of Fig.~\ref{fig1} gives an example of simplest
$1D$ network which can self-sustainedly oscillate. However,
interaction structures of complex networks are in general much more
complex than Fig.~\ref{fig1} which are high dimensional and random
(e.g. Fig.~\ref{fig2}(a)). In Fig.~\ref{fig1} the only topological
loop is right the source loop generating the oscillation. In
practical complex connected networks we usually have $M$ $\gg$ $N$,
and the number of topological loops is large. We therefore propose
an operable and physically meaningful method to reduce the original
random network graphs (as Fig.~\ref{fig2}(a)) to the simple and
instructive TFL patterns of Fig.~\ref{fig1}. The method consists of
the following Complexity Reduction steps:

(a) Find phase-advanced driving interactions for each node.

(b) Find the single dominant interaction among these phase-advanced
driving interactions.

(c) Use all these dominant interactions to unidirectionally link the
network nodes, and draw the dominant phase-advanced path pattern,
which represents the dynamic geodesic of the complex network,
showing the most significant (or say the shortest) driving paths in
the given oscillatory states. This pattern reduce the original
network $G(V,E)$ to a new graph $G(V,E')$ with $E'$ being a subset
of $E$ ($E'\in E$).

  Note, we have $M'=N$ since each node must have one and only one
dominant phase-advanced driving, and thus this graph is nothing but
the $1D$ TFL pattern of Fig.~\ref{fig1}, where the loop is the core
topology for the oscillation serving as the source loop and the
unidirectional links indicate the wave propagation pathways.

  All the above steps rely on only the necessary regulating topology
under the general condition: any nonoscillatory node can oscillate
only if it is driven by phase-advanced interactions from other
nodes. They are widely applicable in diverse fields for
self-sustained oscillations of complex networks of individually
nonoscillatory nodes, independing of node dynamics (excitable or
nonexcitable), coupling forms (active interactions or repressive
interactions, directional or symmetric couplings), and network
structures (small-world networks, purely random networks, or scale
free networks). The particular meanings of $''$advanced phase$''$
and $''$dominant phase-advanced driving$''$ should be properly
defined, according to realistic physical, chemical and biological
interaction mechanisms in each individual system. In the following
sections we will apply this method to particular systems of
oscillatory excitable small-world networks.

\section{Self-sustained target waves in excitable small-world networks}

  We take a two-dimensional ($2D$) B$\ddot{a}$r model \cite{Bar1993} as our example
\begin{subequations}
\label{eq1}
\begin{equation}
\dot{u}_{i,j}=-\frac{1}{\varepsilon}u_{i,j}(u_{i,j}-1)(u_{i,j}-\frac{v_{i,j}+b}{a})+D_{i,j},
\label{eq1a}
\end{equation}
\begin{equation}
\dot{v}_{i,j}=f(u_{i,j})-v_{i,j},
\end{equation}
\begin{displaymath}
D_{i,j}=D_u(u_{i-1,j}+u_{i+1,j}+u_{i,j-1}+u_{i,j+1}-4u_{i,j}).
\end{displaymath}
\end{subequations}
where $f(u_{i,j})=0$ for $u_{i,j} < \frac{1}{3}$;
$f(u_{i,j})=1-6.75u_{i,j}(u_{i,j}-1)^2$ for $\frac{1}{3} \leq
u_{i,j} \leq 1$; and $f(u_{i,j})=1$ for $u_{i,j}
> 1$. The system parameters are kept throughout this paper as  $a=0.84,
b=0.07 ,\varepsilon=0.04$ and $D_u=1.0$ just for the sake that the
local cell follows excitable dynamics. In the present paper we
consider $100 \times 100$ cells in the $2D$ regular lattice with
constant and homogeneous nearest couplings. With this coupling
topology, spiral waves can be easily observed for random initial
conditions. It is well known (and also it is verified in our
simulations) that target waves can never be observed in the
asymptotic states of Eq.~(\ref{eq1}) unless some persistent external
pacings provide the wave sources. Many systems of practical
importance, such as neural systems, have complex coupling structures
where cells can be coupled to each other with both short-range and
long-range couplings. For studying the influence of long-range
couplings (L-RCs) we add an additional coupling term $D^{'}_{i,j}$
to Eq.~(\ref{eq1a}) as
\begin{displaymath}
\dot{u}_{i,j}=-\frac{1}{\varepsilon}u_{i,j}(u_{i,j}-1)(u_{i,j}-\frac{v_{i,j}+b}{a})+D_{i,j}+D^{'}_{i,j},
\end{displaymath}
\begin{equation}
\dot{v}_{i,j}=f(u_{i,j})-v_{i,j},
\label{eq2}
\end{equation}
\begin{displaymath}
D^{'}_{i,j}=\left\{\begin{array}{cc} D_u(u_{i^{'},j^{'}}-u_{i,j}) & for (i,j; i^{'},j^{'})\in \Omega,\\
0, &otherwise,\end{array}\right.
\end{displaymath}
where $\Omega$ is a set of $K$ L-RCs between non-neighbor sites
$(i,j)$ and $(i^{'},j^{'})$ randomly chosen in the $2D$ lattice.
Eq.~(\ref{eq2}) is integrated by the second-order Runge-Kutta scheme
with the time step $\Delta t=0.031$ and the No-Flux boundary
condition is used.

\begin{figure}
\resizebox{14.0cm}{14.0cm}{\includegraphics{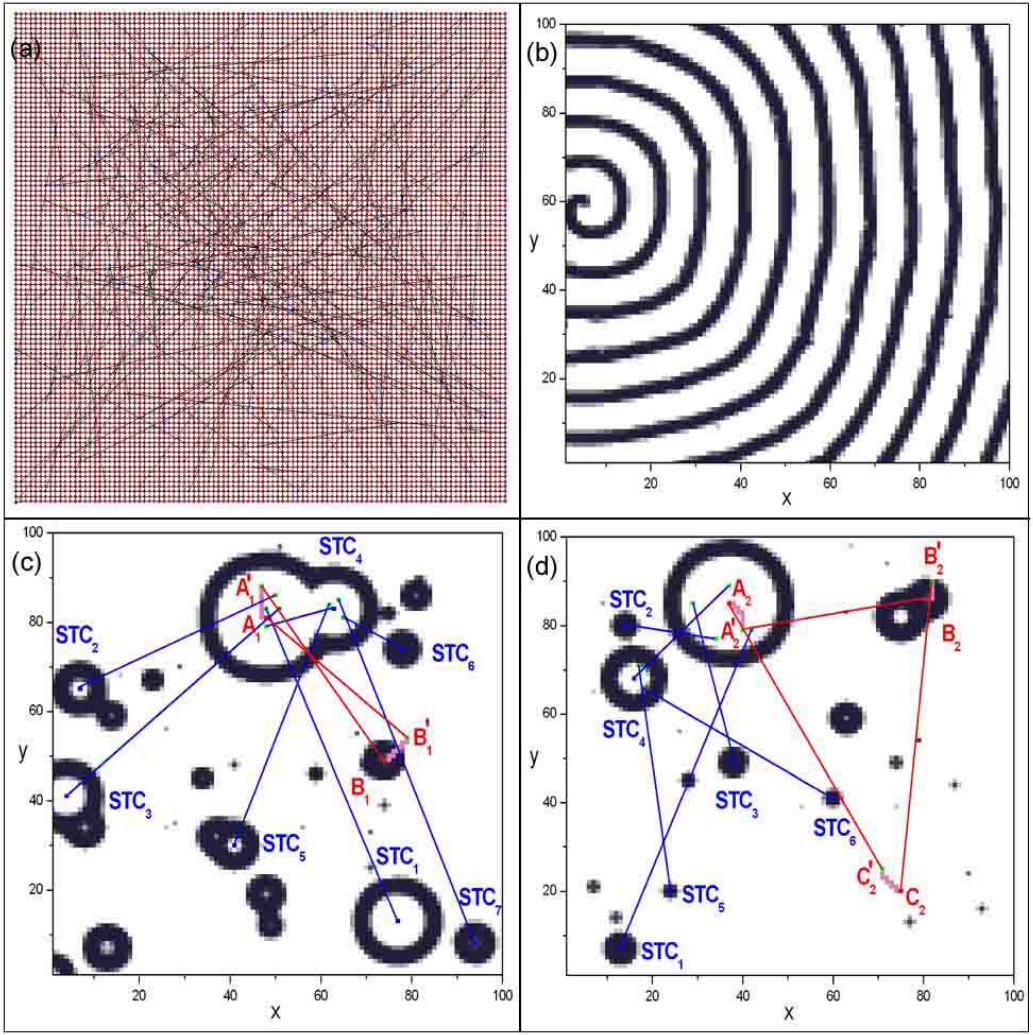}}\quad
\caption{\label{fig2} Coloured online. (a) A small-world network
with 150 random long-range links (note, the tissue has about
$2\times 10^4$ nearest-neighbor links). (b) Spiral waves realized
from a set of initial condition. (c) (d) Different asymptotic target
wave patterns from two different initial conditions. The red nodes
$(A_1,B_1)$ and $(A_2,B_2,C_2)$ denote the centers of the target
waves located in the source loops of Fig.~\ref{fig3}, called source
centers, and the blue nodes $STC_i (i=1,2,\cdots,)$ denote the
sub-target centers driven by the source targets. All the green nodes
denote the nodes connected with the various target centers with
long-range interactions. The source loops found in Fig.~\ref{fig3}
are denoted here $A_1 \xrightarrow{bold \ line} A^{'}_1
\xrightarrow{L-RC} B_1 \xrightarrow{bold \ line} B^{'}_1
\xrightarrow{L-RC} A_1$ in (c); and $A_2 \xrightarrow{bold \ line}
A^{'}_2 \xrightarrow{L-RC} B_2 \xrightarrow{bold \ line} B^{'}_2
\xrightarrow{L-RC} C_2 \xrightarrow{bold \ line} C^{'}_2
\xrightarrow{L-RC} A_2$ in (d).}
\end{figure}

  Now we consider a small-world network of Eq.~(\ref{eq2}) with $K=150$ random L-RCs
which is shown in Fig.~\ref{fig2}(a). We run the system from $100$
sets of random initial conditions and find $3$ realizations for
homogeneous rest state; $63$ for various spiral wave patterns (one
of them is shown in Fig.~\ref{fig2}(b)); and $34$ for self-sustained
target wave patterns (two of them are shown in Figs.~\ref{fig2}(c)
and \ref{fig2}(d)). The mechanism of oscillation of the spiral wave
of Fig.~\ref{fig2}(b) is well understood: the spiral tip plays the
role of oscillation source and waves propagate from the tip to far
away. However, from the randomly distributed groups of target waves
of Figs.~\ref{fig2}(c) and \ref{fig2}(d)  we can hardly say anything
about the mechanism underlying different patterns, for instance,
where the oscillation sources are and how waves propagate from the
sources to the whole tissue. Specifically, in both patterns there
are large numbers of target centers, we may ask which centers are
the true centers (sources) of the oscillations.

  We apply the DPAD method to analyze the oscillation patterns of
Figs.~\ref{fig2}(c) and \ref{fig2}(d). First we should specify the
definitions of phase-advanced and dominant phase-advanced driving
for the sepcific systems of Eq.~(\ref{eq2}). For a given $ith$ cell,
we define the phase-advanced driving as the interactions from all
the neighbors which have $u(t)>u_i(t)$ at time $t_e$ when the $ith$
cell is kicked from the rest state (i.e., $u_i(t)$ crosses the
threshold value $u_e=\frac{b}{a}$ from small value). Among all these
phase-advanced interactions we define the interaction from the node,
which has the largest $u(t)$ at $t_e$, as the dominant driving.
Therefore, each cell is associated with only a single DPAD. These
definitions are clearly demonstrated in Fig.~\ref{fig3}(a) by
studying a node $8250$ of Fig.~\ref{fig2}(c) as an example where
node $8249$ provides the dominant phase-advanced driving to the
given node $8250$, and we can draw an arrowed driving path from
$8249$ to $8250$ in the corresponding TFL pattern of
Fig.~\ref{fig3}(b).

  Now we use this definition to draw the TFL patterns based on
known interaction structure and oscillation data.
Figures.~\ref{fig3}(b) and \ref{fig3}(c) show TFL structures
corresponding to the oscillation patterns of Figs.~\ref{fig2}(c) and
\ref{fig2}(d), respectively. From the TFL patterns all the above
questions can be understood without any ambiguity. First, we have
revealed in Fig.~\ref{fig3}(b) and Fig.~\ref{fig3}(c) the wave
sources \--- 1D source loops (linked by the pink nodes) from the
large number of possible candidates of topological loops in
Fig.~\ref{fig2}(a). In the source loops all red bold arrowed lines
show L-RCs ($A^{'}_1 \longrightarrow B_1$, $B^{'}_1 \longrightarrow
A_1$ in Fig.~\ref{fig3}(b); $A^{'}_2 \longrightarrow B_2$, $B^{'}_2
\longrightarrow C_2$, $C^{'}_2 \longrightarrow A_2$ in
Fig.~\ref{fig3}(c)) and all other arrowed lines come from local
chains. In Figs.~\ref{fig2}(c) and \ref{fig2}(d) all the local
chains in the source loops are shown by bold lines, and the source
centers in each pattern are identified by red disks, such as: double
centers $(A_1,B_1)$ with the source loop $A_1 \xrightarrow{bold \
line} A^{'}_1 \xrightarrow{L-RC} B_1 \xrightarrow{bold \ line}
B^{'}_1 \xrightarrow{L-RC} A_1$ for Fig.~\ref{fig2}(c); and triple
centers $(A_2,B_2,C_2)$ with the source loop $A_2 \xrightarrow{bold
\ line} A^{'}_2 \xrightarrow{L-RC} B_2 \xrightarrow{bold \ line}
B^{'}_2 \xrightarrow{L-RC} C_2 \xrightarrow{bold \ line} C^{'}_2
\xrightarrow{L-RC} A_2$ for Fig.~\ref{fig2}(d).

  Via TFL patterns of Figs.~\ref{fig3}(b) and \ref{fig3}(c) we can not only
understand the problem which target centers are the true centers of
the oscillations (red square nodes), but also understand how waves
propagate from the source centers to all the sub-target centers
($STCs$, shown by blue square nodes), and then produce groups of
target waves successively in the patterns. Taking the sub-target
center $STC_6$ in Fig.~\ref{fig2}(c) as an example, we realize from
the TFL pattern of Fig.~\ref{fig3}(b) that waves start from the
source target center $A_1$ and propagate through the path $A_1
\longrightarrow STC_4 \longrightarrow STC_6$. All these sub-target
centers take their fixed positions in the TFL structures, and the
problems how these centers are driven by upstream nodes (through
L-RCs) and how they drive their downstream nodes (through various
sub-target waves supported by local couplings) are illustrated
clearly.

\section{Control of self-sustained oscillations in excitable complex networks}

  The most interesting point is that we can perform pattern control
and regulation based on the TFL patterns. Our task is to effectively
suppress oscillations. By $''$effective$''$ we mean to change as
small as possible number of couplings. If TFL patterns in
Fig.~\ref{fig3} make sense we expect that removing a single L-RC on
the source loop can suppress the given target waves in the whole
pattern. In Fig.~\ref{fig4}(a) we plot
$<u(t)>=\frac{1}{N^2}\sum^{N=100}_{i,j=1}u_{i,j}(t)$ and
$<v(t)>=\frac{1}{N^2}\sum^{N=100}_{i,j=1}v_{i,j}(t)$ vs $t$, and
show that $<u(t)>$ and $<v(t)>$ damps to zero after a single L-RC
$B^{'}_1A_1$ of Fig.~\ref{fig2}(c) is discarded. A snapshot of
pattern evolution after discarding $B^{'}_1A_1$ is presented in
Fig.~\ref{fig4}(b) where the system is approaching to the
homogeneous rest state. The damping process after discarding L-RC
$B^{'}_1A_1$ can be well explained based on the TFL pattern of
Fig.~\ref{fig3}(b). When we remove $B^{'}_1A_1$, the source loop
$A_1 \longrightarrow A^{'}_1 \longrightarrow B_1 \longrightarrow
B^{'}_1 \longrightarrow A_1$ breaks, and thus the source centers
$A_1$, $B_1$ no longer emit waves, and the target waves from centers
$A_1$, $B_1$ first damp. Without excitations from these sources, all
other target centers cease to emit waves successively due to the
successive annihilations of their driving waves. Finally, the whole
pattern with all waves generated by a large numbers of target
centers evolves to the homogeneous rest state. Similar control
effects are observed for other self-sustained target patterns. When
any single long-range link in the source loop is discarded, the
given oscillation of the whole system is destroyed completely, and
the system either evolves to the homogeneous rest state, or develops
into an entirely different oscillatory structure due to the
reconstruction of a different new source loop. It is really striking
that by discarding only a single L-RC link among $2\times 10^4$
local couplings and $150$ L-RCs we can essentially change the
dynamics of the whole pattern formation. This can never happen for
spiral waves.

  It is emphasized that the self-sustained target waves are robust
against random changes of the coupling structure. For instance, the
main structure of pattern Fig.~\ref{fig2}(c) is not considerably
changed if we discard all L-RCs except the 9 relevant L-RCs shown in
Fig.~\ref{fig2}(c). A snapshot of pattern after discarding such 141
L-RCs are shown in Fig.~\ref{fig4}(c). Fig.~\ref{fig4}(d) shows the
asymptotic pattern when we discard all 148 L-RCs in
Fig.~\ref{fig2}(c) except the $2$ L-RCs $A^{'}_1B_1$ and
$B^{'}_1A_1$ in the source loop. The oscillation continues just with
these two long-range links. The comparison of Fig.~\ref{fig4}(d)
(discarding 148 L-RCs and keeping only the two L-RCs in the source
loop) with Fig.~\ref{fig4}(b) (discarding only a single L-RC in the
source loop) is really striking.

\section{Robustness of Applications of the DPAD method}

  In the above applications of the DPAD method, we required full and
precise knowledge on the interaction structures of networks and the
variable data of oscillatory nodes. In realistic cases, the
information is often not complete, it is thus necessary to consider
the robustness of our approach in these practical situations.

  First, in practical experiments the measured data are often not
precise, i.e., we should consider noisy data. In these cases, our
method has an advantage of robustness. The validity of the DPAD
method does not rely on precise values of data measurements, but
depend on the topological orders of data, such as phase-advanced or
phase-delayed interactions and dominant or nondominant drivings.
These topological orders can often be correctly explored when the
data measurements have certain small errors. Moreover, for periodic
oscillations we may reduce some random measurement errors by
averaging the data in sufficiently long intervals by making optimal
tradeoff between the precision and the measurement expense. Those
treatments are applicable when the errors are relative small. If the
errors are large, the methods may no longer work.

  Now we focus on the second and the most important problem: the
robustness of the DPAD method against the incompleteness of
information. Another promising advantage of our method is that the
pathways of TFL patterns are drawn \emph{based on local knowledge},
i.e., the driving path of a given node is determined by the
oscillation data of this node and few nodes in direct interactions.
Lack of information around certain node does not affect the
applications of the DPAD method to nodes far away. In these cases it
is possible to draw incomplete TFL patterns from incomplete
knowledge, and these unperfect TFL patterns may still provide useful
and instructive guidance in analyzing the oscillations. Here we only
consider a relatively simple situation of information lacks \---
incomplete interaction structure.

 We again consider the state of Fig.~\ref{fig2}(c), and apply
the DPAD method with incomplete interaction knowledge. Specifically,
we draw the TFL pattern under the condition that 50 L-RCs randomly
chosen unknown. One of these TFL patterns is shown in
Fig.~\ref{fig5}(a). Comparing Fig.~\ref{fig5}(a) with
Fig.~\ref{fig3}(b) we find while Fig.~\ref{fig3}(b) has a single
connected network, Fig.~\ref{fig5}(a) has at most 51 clusters
disconnected from each other (only 4 clusters are shown in
Fig.~\ref{fig5}(a)). In the incomplete TFL pattern we observe some
tree-like networks each having a source node controlling the
downstream nodes. We also find a cluster having a TFL loop which is
exactly the same as the source loop of Fig.~\ref{fig3}(b). It is
obvious that Fig.~\ref{fig5}(a) keeps the essential structure of the
complete TFL shown in Fig.~\ref{fig3}(b). In particular, from
Fig.~\ref{fig5}(a) we see clearly the oscillation source and we can
locate the two key L-RCs in the source loop which control the whole
oscillatory state. In Fig.~\ref{fig5}(b) we present another TFL
pattern for a different set of unknown 50 long-range interactions,
where a L-RC on the source loop of Fig.~\ref{fig3}(b) is,
unfortunately, chosen to be unknown. Now we observe no loop but at
most 50 tree clusters ($4$ clusters shown in Fig.~\ref{fig5}(b)),
each is driven by a source node. And the source nodes themselves are
not driven by any other node in the incomplete TFL. It becomes
difficult to locate the source loop without some additional
information. Nevertheless, the incomplete TFL pattern of
Fig.~\ref{fig5}(b) can still show rich driving paths of the state
Fig.~\ref{fig2}(c) and it may serve as an excellent guidance for
analyzing the oscillation organization. The following understanding
is of great help in estimating the unknown L-RCs of the source
nodes. (\romannumeral 1) Each source node must be driven by a L-RC
phase-advanced. (\romannumeral 2) In TFL pattern the phase
differences between all pairs of target centers and their
corresponding upstream driving nodes are approximately the same
(they are about $1.91$ $rad$ in our system as shown in
Fig.~\ref{fig5}(c)). This phase difference can be used for seeking
the upstream driving nodes of the unknown L-RCs. (\romannumeral 3)
The $''$source$''$ node controlling the largest tree cluster of
Fig.~\ref{fig5}(b) has large probability to be a key node in the
source loop (this conclusion is confirmed by all 10 test, using
different sets of unknown L-RCs). These conclusions, which may be
popular for self-sustained target waves in small-world excitable
networks, can considerably reduce the difficulty in recovering the
complete TFL. If we have some additional (still not complete)
information which may be model dependent, the above conclusions may
be used for predicting the missing long-range driving links of the
source nodes and recovering the complete TFL structure. For
instance, in our system all the periodic orbits of upstream driving
nodes of L-RCs have similar characteristic which is considerably
different from normal nodes (see Fig.~\ref{fig5}(c) for an example).
Jointly using this particular feature and the above three general
conclusions we can explore the missing L-RC drivings of source loop
and all other missing L-RCs of source nodes with extremely high
probability. For instance, in 10 random samples we successfully
predict all the missing L-RCs of source nodes and recover the
original main TFL structure without any error, of which the patching
from Fig.~\ref{fig5}(b) is shown in Fig.~\ref{fig5}(d).

Furthermore, we have tested many other cases with different system
sizes, different numbers of existing L-RCs and different unknown
L-RCs. We found that the above method of exploring unknown L-RCs
works fairly well, and we can correctly reveal TFL loops with high
probability. However, the method is not always successful. It may
fail when different source nodes are linked by L-RCs having similar
time series. The recover operations will be even more difficult when
both node data and interaction structure are incomplete. We will go
further into these cases in our future study.
\section{Conclusion}

  In conclusion we have studied periodic self-sustained oscillations
of target waves in excitable small-world networks. Based on
interaction structures and oscillation data, a novel method of
dominant phase-advanced driving (DPAD) path is proposed to study
self-sustained oscillations in complex networks of nonoscillatory
nodes and tree from loop (TFL) patterns can be drawn for oscillatory
states. For oscillatory excitable small-world networks these TFL
patterns show clearly 1D loops serving as oscillation sources and
illustrate how waves propagate from the source loops to the whole
tissues via various local and long-range couplings of small-world
networks. More interestingly, with the instructive information of
TFL patterns we can control and regulate oscillatory patterns of
small-world networks with extremely high efficiency.
Self-sustainedly oscillatory complex networks of excitable cells
exist significantly in wide fields, such as neural systems. Both the
DPAD method and TFL structures are thus expected to be of broad
interest in crossing fields of nonlinear dynamics, complex networks
and their practical applications. In the present paper we consider
only periodic oscillations of fixed complex networks. The extensions
to nonperiodic oscillations (e.g., chaotic) and to the cases of
noise and parameter-variation-driven oscillations, which occur
popularly in practical situations, are future interesting tasks.

\begin{acknowledgments}
  This work was supported by the National Natural Science Foundation
of China under Grant No. 10675020 and by the National Basic Research
Program of China (973 Program) under Grant No. 2007CB814800.
\end{acknowledgments}

\newpage 
\bibliography{Ref}

\begin{thebibliography}{22}
\bibitem[{Pha()}]{Phaseadvanced}
\bibinfo{note}{For a given node (say, node A), we define all interactions
  favorable to activating node A from the rest state as phase-advanced
  interactions, and among the phase-advanced interactions we define the
  interaction providing the most significant activating contribution as the
  dominant phase-advanced driving (DPAD)}.
\end{thebibliography}

\newpage

\newpage

\newpage

\begin{figure}
\resizebox{14.0cm}{14.0cm}{\includegraphics{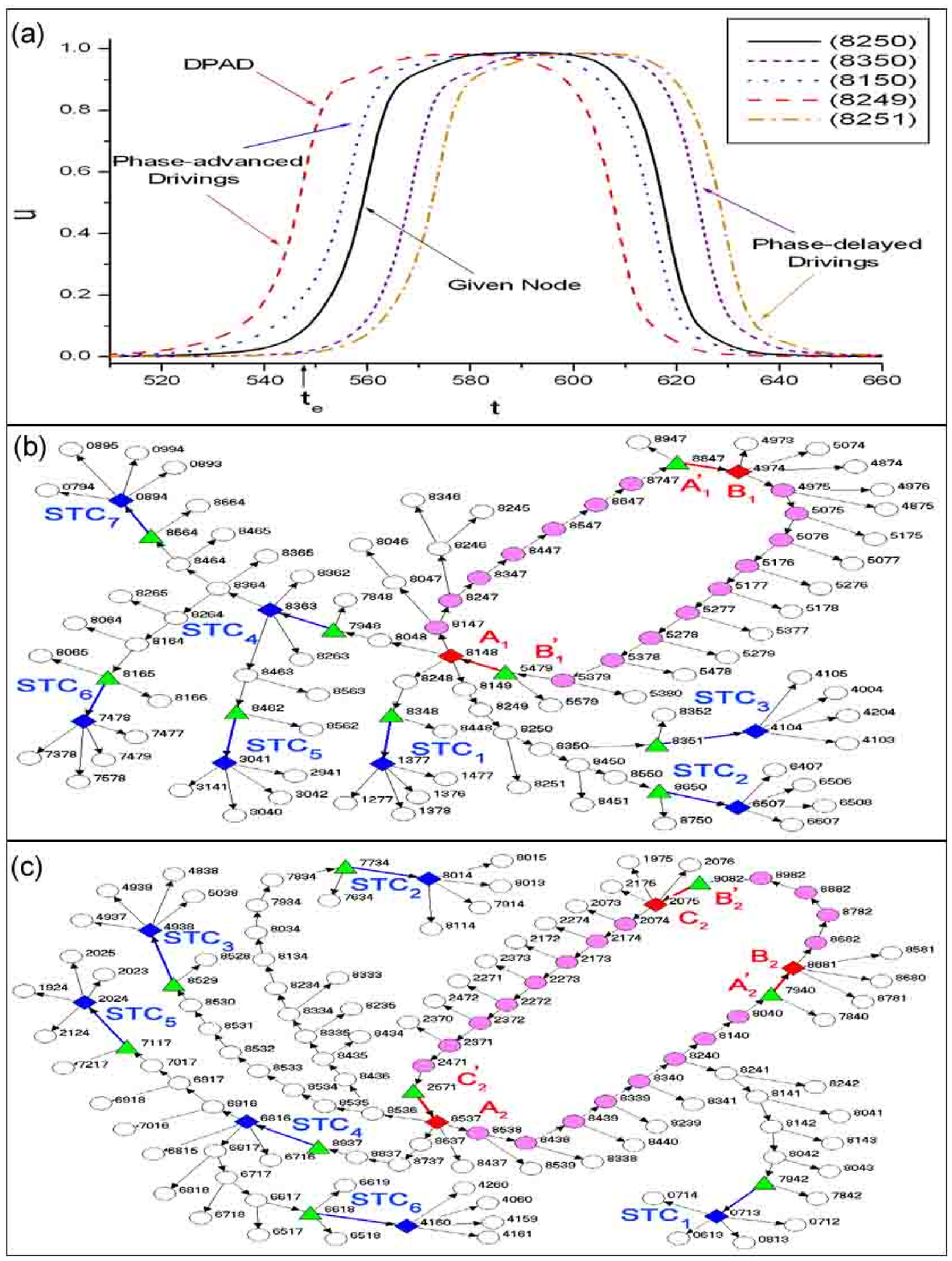}}\quad
\end{figure}

\newpage

\begin{figure}
\caption{\label{fig3} Coloured online. (a) Demonstration of the
phase-advanced drivings and dominant phase-advanced driving of a
given node $8250$ for the state Fig.~\ref{fig2}(c). Black solid
curve shows the $u(t)$ signal of the given node $8250$. Other four
coloured curves are the $u(t)$ signals of nodes interacting with
node $8250$. Jumping time $t_e$ is marked by the arrowed line. It is
obvious that signals $8249$ (red dash line) and $8150$ (blue dot
line) represent phase-advanced drivings, and their couplings help to
excite the given node at $t_e$. And the node $8249$ (red dash line)
signal provides the most significant contribution in exciting the
given node, and is identified as dominant phase-advanced driving.
Signals $8350$ (purple short dash line) and $8251$ (yellow dash dot
line) are phase-delayed interactions, and they provide no (or,
precisely, play negative) contributions to kick the given node. (b)
(c) TFL structures (partial nodes and DPADs are presented)
corresponding to patterns Figs.~\ref{fig2}(c) and \ref{fig2}(d),
respectively. All subscripts indicate the node positions, e.g.,
$8250$ representing the node index $(i,j)$ with $i=50$, $j=82$. All
red (arrowed to the red square source target centers) and blue
(arrowed to the blue square sub-target centers) bold lines denote
long-range interactions. From these TFL patterns we can identify the
following information: (i) DPAD loops (linked by pink nodes) as the
oscillation sources; (ii) Oscillation centers (red square nodes in
source loops), i.e., $(A_1,B_1)$ for (b) and $(A_2,B_2,C_2)$ for (c)
from which all waves are generated; (iii) Successive driving
sequences representing wave propagation pathways.}
\end{figure}

\newpage

\begin{figure}
\resizebox{14.0cm}{14.0cm}{\includegraphics{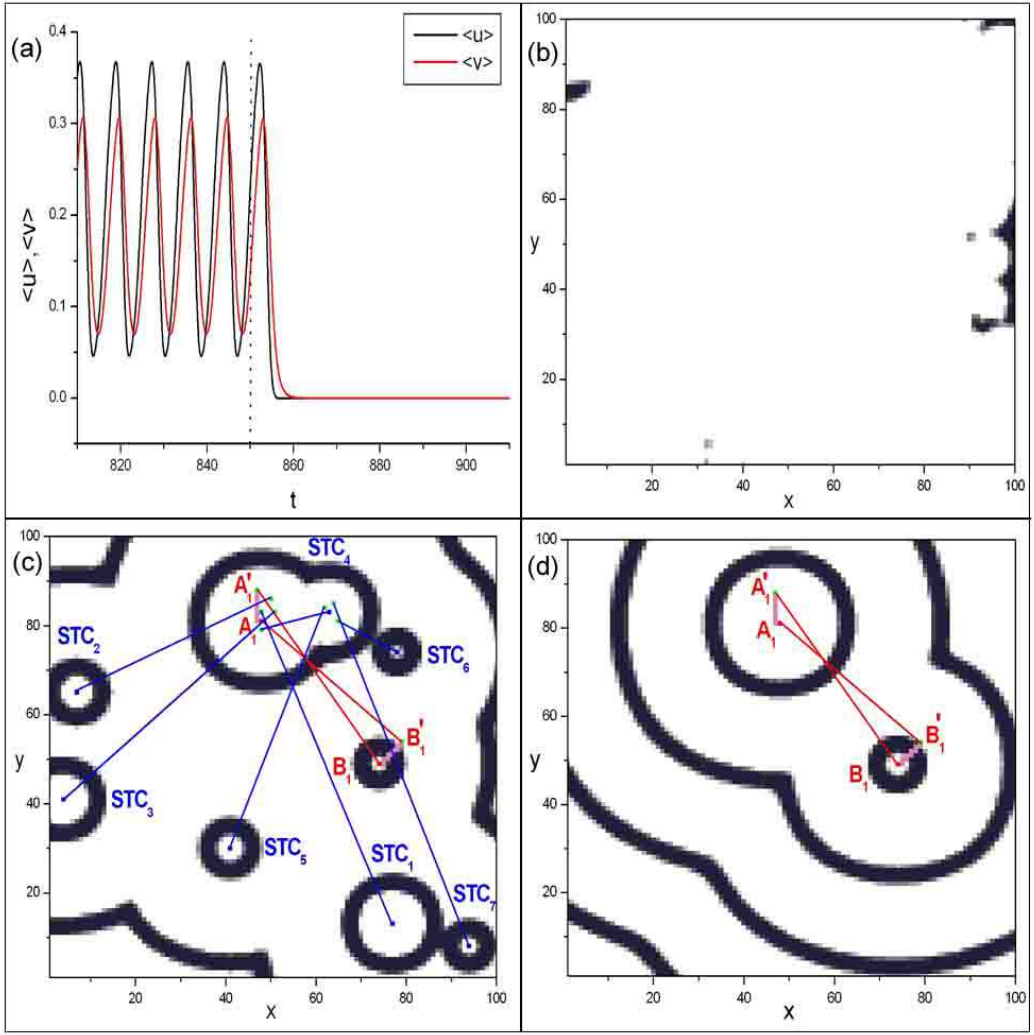}}\quad
\caption{\label{fig4} Coloured online. Oscillation regulation of
target waves by discarding some interacting links of Fig.~\ref{fig2}
(c). (a) Trajectories of
$<u(t)>\frac{1}{N^2}\sum^{N=100}_{i,j=1}u_{i,j}(t)$ and
$<v(t)>\frac{1}{N^2}\sum^{N=100}_{i,j=1}v_{i,j}(t)$ with a single
long-range link $B^{'}_1A_1$ of pattern Fig.~\ref{fig2}(c) discarded
at $t=850$ (indicated by the dot line). (b) A snapshot of pattern
taken at $t=855$ evolved from pattern Fig.~\ref{fig2}(c) after the
L-RC $B^{'}_1A_1$ discarded. (c) (d) Snapshots of the asymptotic
patterns taken at $t=1500$ with 141 long-rang links ((c)) and 148
long-rang links ((d)) discarded at $t=850$, respectively. All the
remained long-range couplings are shown in (c) and (d).}
\end{figure}

\newpage

\begin{figure}
\resizebox{14.0cm}{14.0cm}{\includegraphics{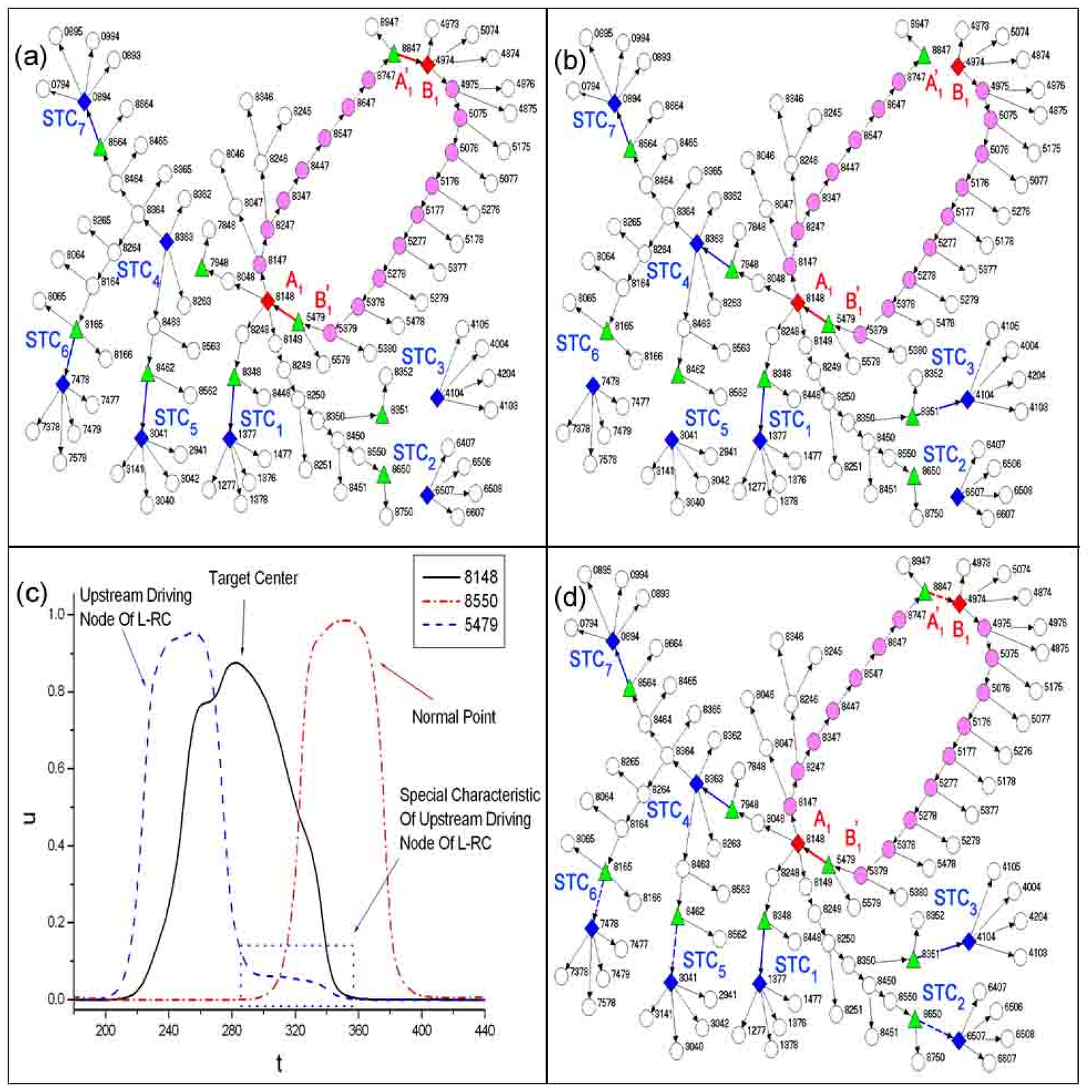}}\quad
\end{figure}

\newpage

\begin{figure}
\caption{\label{fig5} Coloured online. DPAD paths of state
Fig.~\ref{fig2}(c) with incomplete information. (a) TFL pattern with
50 L-RCs, randomly chosen, unknown. The pattern keeps the essential
structure and exactly the same source loop of Fig.~\ref{fig3}(b).
(b) The same as (a) with different set of 50 L-RCs unknown. Now the
source loop of Fig.~\ref{fig3}(b) cannot be explored directly.
However, this incomplete DPAD pattern is still very useful in
exploring the driving structure of state Fig.~\ref{fig2}(c). (c)
Demonstration of the characteristics of periodic orbits of various
nodes. Black solid, blue dash and red dash dot curves show $u(t)$
signals of a given target center 8148, a node 5479 driving the
target center 8148 via a L-RC, and a node 8550 without the L-RC
driving (called normal node), respectively. In (c) the L-RC driving
node (blue dash curve) has the characteristic of slightly hampered
tail (see the small dot frame), clearly distinguishing from the
normal node (red dash dot curve with perfectly smooth tail). This
distinction has been verified to be common for all nodes testified.
(d) By applying the three conclusions in Sec. \Rmnum{5} and the
characteristic distinctions shown in Figs.~\ref{fig5}(c) we can
supplement all the missing long-range driving links of the source
nodes in Figs.~\ref{fig5}(a) and \ref{fig5}(b) (see dot arrows in
(d) for patching Fig.~\ref{fig5}(b)). Now the TFL pattern patched
from Fig.~\ref{fig5}(b) shows exactly the same self-organized
oscillation structure as Fig.~\ref{fig3}(b).}
\end{figure}

\end{document}